\documentstyle[12pt]{article}
\def\href#1#2{{#2}}
\begin{document}
\begin{titlepage}
\begin{flushright}
quant-ph/9503015 \\
THEP-95-3 \\
March 1995
\end{flushright}
\vspace{0.2cm}
\begin{center}
\LARGE
HyperDiamond Feynman Checkerboard \\
in 4-dimensional Spacetime \\
\vspace{1cm}
\normalsize
Frank D. (Tony) Smith, Jr. \\
\footnotesize
e-mail: gt0109e@prism.gatech.edu \\
and fsmith@pinet.aip.org  \\
P. O. Box for snail-mail: \\
P. O. Box 430, Cartersville, Georgia 30120 USA \\
\href{http://www.gatech.edu/tsmith/home.html}{WWW
URL http://www.gatech.edu/tsmith/home.html} \\
\vspace{12pt}
School of Physics  \\
Georgia Institute of Technology \\
Atlanta, Georgia 30332 \\
\vspace{0.2cm}
\end{center}
\normalsize
\begin{abstract}
A generalized Feynman Checkerboard model is
constructed using a 4-dimensional HyperDiamond lattice.
The resulting phenomenological model is the
$D_{4}-D_{5}-E_{6}$ model described in
\href{http://xxx.lanl.gov/abs/hep-ph/9501252}{hep-ph/9501252}
and
\href{http://xxx.lanl.gov/abs/quant-ph/9503009}{quant-ph/9503009}.
\end{abstract}
\vspace{0.2cm}
\normalsize
\footnoterule
\noindent
\footnotesize
\copyright 1995 Frank D. (Tony) Smith, Jr., Atlanta, Georgia USA
\end{titlepage}
\newpage
\setcounter{footnote}{0}
\setcounter{equation}{0}

\tableofcontents

\newpage

\section{Introduction.}

The 1994 Georgia Tech Ph. D. thesis of Michael Gibbs under
David Finkelstein \cite{GIB} constructed a discrete 4-dimensional
spacetime HyperDiamond lattice, here denoted a $4HD$ lattice,
in the course of building a physics model.
\vspace{12pt}

The 1992 MIT Ph. D. thesis of Hrvoje Hrgovcic under
Tommasso Toffoli \cite{HRG} constructed a discrete 4-dimensional
spacetime Minkowski lattice in the course of building
a simulation of solutions to the Dirac equation and other equations.
\newline
Hrgovcic's lattice is closely related to the $4HD$ lattice.
\vspace{12pt}

Both the Gibbs model and the Hrgovcic simulation differ in
significant ways from the $D_{4}-D_{5}-E_{6}$ model constructed
from $3 \times 3$ octonion matrices
\newline
in
\href{http://xxx.lanl.gov/abs/hep-ph/9501252}{hep-ph/9501252}
and
\href{http://xxx.lanl.gov/abs/quant-ph/9503009}{quant-ph/9503009}.
\vspace{12pt}

However, both theses have been very helpful with respect to
this paper, which generalizes the 2-dimensional Feynman
checkerboard \cite{FEY1, FEY2} to a physically realistic
4-dimensional spacetime.
\vspace{12pt}

In this 4-dimensional $4HD$ lattice generalized Feynman
\newline
HyperDiamond checkerboard, the properties of the particles that move
\newline
around on the checkerboard are determined by the physically
\newline
realistic $D_{4}-D_{5}-E_{6}$ model
constructed from $3 \times 3$ octonion matrices
\newline
in
\href{http://xxx.lanl.gov/abs/hep-ph/9501252}{hep-ph/9501252}
and
\href{http://xxx.lanl.gov/abs/quant-ph/9503009}{quant-ph/9503009}.
\vspace{12pt}

The 4-dimensional HyperDiamond $4HD$ checkerboard is related to
\newline
the 8-dimensional HyperDiamond $8HD$ = $E_{8}$ in the same way
\newline
the 4-dimensional physical associative spacetime is related to
\newline
the 8-dimensional octonionic spacetime in
the $D_{4}-D_{5}-E_{6}$ model,
\newline
so the HyperDiamond $4HD$ Feynman checkerboard is fundamentally
\newline
the discrete version of the $D_{4}-D_{5}-E_{6}$ model.
\vspace{12pt}

For more about octonions and lattices, see the works of
Goeffrey Dixon \cite{DIX4, DIX5, DIX6, DIX7}.

\newpage

\section{Feynman Checkerboards.}

The 2-dimensional Feynman checkerboard \cite{FEY1, FEY2}
is a notably successful and useful representation of
the Dirac equation in 2-dimensional spacetime.
\vspace{12pt}

To build a Feynman 2-dimensional checkerboard,
\newline
start with a 2-dimensional Diamond checkerboard with
\newline
two future lightcone links and two past lightcone
links at each vertex.
\vspace{12pt}

The future lightcone then looks like
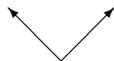
\vspace{12pt}

\begin{picture}(50,50)

\put(25,10){\vector(1,1){20}}
\put(25,10){\vector(-1,1){20}}

\end{picture}

\vspace{12pt}

If the 2-dimensional Feynman checkerboard is
\newline
coordinatized by the complex plane $\bf{C}$:
\newline
the real axis $1$ is identified with the time axis $t$;
\newline
the imaginary axis $i$ is identified with the space axis $x$; and
 \newline
the two future lightcone links are
$(1/\sqrt{2})(1 + i)$ and $(1/\sqrt{2})(1 - i)$.
\vspace{12pt}

In cylindrical coordinates $t,r$ with $r^{2} = x^{2}$,
\newline
the Euclidian metric is $t^{2} + r^{2}$ = $t^{2} + x^{2}$ and
\newline
the Wick-Rotated Minkowski metric with speed of light $c$ is
\newline
 $(ct)^{2} - r^{2}$ = $(ct)^{2} - x^{2}$.
\vspace{12pt}

For the future lightcone links to lie on
\newline
the 2-dimensional Minkowski lightcone, $c = 1$.
\vspace{12pt}

Either link is taken into the other link
by complex multiplication by $\pm i$.
\vspace{12pt}

Now, consider a path in the Feynman checkerboard.
\newline
At a given vertex in the path, denote the future lightcone link in
\newline
the same direction as the past path link by $1$, and
\newline
the future lightcone link in
the (only possible) changed direction by $ i$.
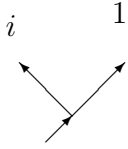
\vspace{12pt}

\begin{picture}(50,50)

\put(15,0){\vector(1,1){10}}
\put(25,10){\vector(1,1){20}}
\put(25,10){\vector(-1,1){20}}
\put(40,45){$1$}
\put(0,40){$ i$}

\end{picture}

\vspace{12pt}

The Feynman checkerboard rule is that
\newline
if the future step at a vertex point of a given path
is in a different direction
\newline
from the immediately preceding step from the past,
\newline
then the path at the point of change gets a weight
of $ -i m \epsilon$,
\newline
where $m$ is the mass
(only massive particles can change directions),and
\newline
$\epsilon$ is the length of a path segment.
\vspace{12pt}

Here I have used the Gersch \cite{GER} convention
\newline
of weighting each turn by $ -im \epsilon$
\newline
rather than the Feynman \cite{FEY1, FEY2} convention
\newline
of weighting by $ +im \epsilon$, because Gersch's
convention gives a better nonrelativistic limit
in the isomorphic 2-dimensional Ising model \cite{GER}.
\vspace{12pt}

HOW SHOULD THIS BE GENERALIZED
\newline
TO HIGHER DIMENSIONS?
\vspace{12pt}

The 2-dim future light-cone is the 0-sphere $S^{2-2} = S^{0} =
\{ i, 1 \}$ ,
\vspace{12pt}

with $1$ representing a path step to the future in
\newline
the same direction as the path step from the past, and
\vspace{12pt}

$ i$ representing a path step to the future in a
\newline
(only 1 in the 2-dimensional Feynman checkerboard lattice)
\newline
different direction from the path step from the past.
 \vspace{12pt}

The 2-dimensional Feynman checkerboard lattice spacetime can be
\newline
represented by the complex numbers $\bf{C}$,
\newline
with $1,i$ representing the two future lightcone directions and
\newline
$-1,-i$ representing the two past lightcone directions.
\vspace{12pt}

Consider a given path in
\newline
the Feynman checkerboard lattice 2-dimensional spacetime.
\vspace{12pt}

At any given vertex on the path in the lattice 2-dimensional
spacetime,
\newline
the future lightcone direction representing the
continuation of the path
\newline
in the same direction can be represented by $1$, and
\newline
the future lightcone direction representing the (only 1 possible)
\newline
change of direction can be represented by $ i$ since either
\newline
of the 2 future lightcone directions can be taken into the other
\newline
by multiplication by $\pm i$,
\newline
$+$ for a left turn and $-$ for a right turn.
\vspace{12pt}

If the path does change direction at the vertex,
\newline
then the path at the point of change gets a weight of
$ -im \epsilon$,
\newline
where $i$ is the complex imaginary,
\newline
$m$ is the mass (only massive particles can change directions),
and
\newline
$\epsilon$ is the timelike length of a path segment,
where the 2-dimensional speed of light is taken to be $1$.
\vspace{12pt}

Here I have used the Gersch \cite{GER} convention
\newline
of weighting each turn by $ -im \epsilon$
\newline
rather than the Feynman \cite{FEY1, FEY2} convention
\newline
of weighting by $ +im \epsilon$, because Gersch's
convention gives a better nonrelativistic limit
in the isomorphic 2-dimensional Ising model \cite{GER}.
\vspace{12pt}

For a given path, let
\newline
$C$ be the total number of direction changes, and
\newline
$c$ be the $c$th change of direction, and
\newline
$ i$ be the complex imaginary  representing
the $c$th change of direction.
\vspace{12pt}

$C$ can be no greater than the timelike checkerboard distance $D$
\newline
between the initial and final points.
\vspace{12pt}

The total weight for the given path is then
\vspace{12pt}

\begin{equation}
\prod_{0 \leq c \leq C} -i m \epsilon =
(m \epsilon)^{C} (\prod_{0 \leq c \leq C} -i) =
(-im \epsilon )^{C}
\end{equation}

\vspace{12pt}

The product is a vector in the direction $\pm 1$ or $\pm i$.
\vspace{12pt}

Let $N(C)$ be the number of paths with $C$ changes in direction.
\vspace{12pt}

The propagator amplitude for the particle to go from
the initial vertex to the final vertex is the sum over all
paths of the weights, that is the path integral sum
over all weighted paths:
\vspace{12pt}

\begin{equation}
\sum_{0 \leq C \leq D} N(C) (-im \epsilon )^{C}
\end{equation}

\vspace{12pt}

The propagator phase is the angle between
\newline
the amplitude vector in the complex plane and the complex real axis.
\vspace{12pt}

Conventional attempts to generalize the Feynman checkerboard from
\newline
2-dimensional spacetime to $k$-dimensional spacetime are based on
\newline
the fact that the 2-dimensional future light-cone directions are
\newline
the 0-sphere $S^{2-2} = S^{0} = \{ i,1 \}$.
\vspace{12pt}

The $k$-dimensional continuous spacetime lightcone directions are
\newline
the $(k-2)$-sphere $S^{k-2}$.
\vspace{12pt}

In 4-dimensional continuous spacetime, the lightcone directions
are $S^{2}$.
\vspace{12pt}

Instead of looking for a 4-dimensional lattice spacetime, Feynman
\newline
and other generalizers went from discrete $S^{0}$
to continuous $S^{2}$
\newline
for lightcone directions, and then tried to construct a weighting
\newline
using changes of directions as rotations
in the continuous $S^{2}$, and
\newline
never (as far as I know) got any generalization that worked.
\vspace{12pt}

\newpage

THE $D_{4}-D_{5}-E_{6}$ MODEL $4HD$ HYPERDIAMOND
\newline
GENERALIZATION HAS DISCRETE LIGHTCONE DIRECTIONS.
\vspace{12pt}

If the 4-dimensional Feynman checkerboard is coordinatized by
\newline
the quaternions $\bf{Q}$:
\newline
the real axis $1$ is identified with the time axis $t$;
\newline
the imaginary axes $i,j,k$ are identified with the space
axes $x,y,z$; and
\newline
the four future lightcone links are
\newline
$(1/2)(1+i+j+k)$,
\newline
$(1/2)(1+i-j-k)$,
\newline
$(1/2)(1-i+j-k)$, and
\newline
$(1/2)(1-i-j+k)$.
\vspace{12pt}

In cylindrical coordinates $t,r$
\newline
with $r^{2} = x^{2}+y^{2}+z^{2}$,
\newline
the Euclidian metric is $t^{2} + r^{2}$ = $t^{2} +
x^{2}+y^{2}+z^{2}$ and
\newline
the Wick-Rotated Minkowski metric with speed of light $c$ is
\newline
 $(ct)^{2} - r^{2}$ = $(ct)^{2} - x^{2} -y^{2} -z^{2}$.
\vspace{12pt}

For the future lightcone links to lie on
\newline
the 4-dimensional Minkowski lightcone, $c = \sqrt{3}$.
\vspace{12pt}

Any future lightcone link is taken into any other future lightcone
\newline
link by quaternion multiplication by $\pm i$,  $\pm j$, or  $\pm k$.
\vspace{12pt}

For a given vertex on a given path,
\newline
continuation in the same
direction can be represented by the link $1$, and
\newline
changing direction can be represented by the
\newline
imaginary quaternion $\pm i,\pm j,\pm k$ corresponding to
\newline
the link transformation that makes the change of direction.
\vspace{12pt}

Therefore, at a vertex where a path changes direction,
\newline
a path can be weighted by quaternion imaginaries
\newline
just as it
is weighted by the complex imaginary in the 2-dimensional case.
\vspace{12pt}

If the path does change direction at a vertex, then
\newline
the path at the point of change gets a weight of
$-im \epsilon$, $-jm \epsilon$, or $-km \epsilon$
\newline
where $i,j,k$ is the quaternion imaginary representing
the change of direction,
\newline
$m$ is the mass (only massive particles can change directions),
and
\newline
$\sqrt{3} \epsilon$ is the timelike length of a path segment,
\newline
where the 4-dimensional speed of light is taken to be $\sqrt{3}$.
\vspace{12pt}

For a given path,
\newline
let $C$ be the total number of direction changes,
\newline
$c$ be the $c$th change of direction, and
\newline
$e_{c}$ be the quaternion imaginary $i,j,k$ representing
the $c$th change of direction.
\vspace{12pt}

$C$ can be no greater than the timelike checkerboard distance $D$
\newline
between the initial and final points.
\vspace{12pt}

The total weight for the given path is then
\vspace{12pt}

\begin{equation}
\prod_{0 \leq c \leq C} -e_{c} m \sqrt{3} \epsilon =
(m \sqrt{3} \epsilon)^{C} (\prod_{0 \leq c \leq C} -e_{c})
\end{equation}

\vspace{12pt}

Note that since the quaternions are not commutative,
\newline
the product must be taken in the correct order.
\vspace{12pt}

The product is a vector in the direction $\pm 1$,
$\pm i$, $\pm j$, or $\pm k$.
and
\vspace{12pt}

Let $N(C)$ be the number of paths with $C$ changes in direction.
\vspace{12pt}

The propagator amplitude for the particle to go
\newline
from the initial vertex to the final vertex is
\newline
the sum over all paths of the weights,
\newline
that is the path integral sum over all weighted paths:
\vspace{12pt}

\begin{equation}
\sum_{0 \leq C \leq D} N(C) (m \sqrt{3} \epsilon)^{C}
(\prod_{0 \leq c \leq C} -e_{c})
\end{equation}

\vspace{12pt}

The propagator phase is the angle between
\newline
the amplitude vector in quaternionic 4-space and
\newline
the quaternionic real axis.
\vspace{12pt}

The plane in quaternionic 4-space defined by
\newline
the amplitude vector and the quaternionic real axis
\newline
can be regarded as the complex plane of the propagator phase.
\vspace{12pt}

\newpage

\section{HyperDiamond Lattices.}

The name "HyperDiamond" was first used by David Finkelstein
in our discussions of these structures.
\vspace{12pt}

n-dimensional HyperDiamond structures $nHD$ are
\newline
constructed from $D_{n}$ lattices.
\vspace{12pt}

An n-dimensional HyperDiamond structures $nHD$ is a lattice
\newline
if and only if $n$ is even.
\newline
If $n$ is odd, the $nHD$ structure is
\newline
only a "packing", not a "lattice", because a nearest neighbor link
\newline
from an origin vertex to a destination vertex
\newline
cannot be extended in the same direction
\newline
to get another
nearest neighbor link.
\vspace{12pt}

n-dimensional HyperDiamond structures $nHD$ are
\newline
constructed from $D_{n}$ lattices.
\vspace{12pt}

The lattices of type $D_{n}$ are n-dimensional checkerboard
lattices, that is,
\newline
the alternate vertices of a ${\bf{Z}}^{n}$ hypercubic lattice.
\vspace{12pt}
A general reference on lattices is Conway and Sloane \cite{CON}.
\vspace{12pt}

For the n-dimensional HyperDiamond lattice construction
from $D_{n}$,
\newline
Conway and Sloane use an
n-dimensional glue vector $[1] = (0.5, ..., 0.5)$
\newline
(with n $0.5$'s).
\vspace{12pt}

Consider the 3-dimensional structure $3HD$.
\vspace{12pt}

Start with $D_{3}$, the fcc close packing in 3-space.
\vspace{12pt}

Make a second $D_{3}$ shifted by the glue vector $(0.5, 0.5, 0.5)$.
\vspace{12pt}

Then form the union $D_{3} \cup  ([1] + D_{3})$.
\vspace{12pt}

That is a 3-dimensional Diamond crystal structure,
\newline
the familiar 3-dimensional thing
\newline
for which HyperDiamond lattices are named.
\vspace{12pt}

\newpage

\subsection{8-dimensional HyperDiamond Lattice.}

When you construct an 8-dimensional HyperDiamond $8HD$ lattice,
\newline
you get $D_{8} \cup  ([1] + D_{8})$ = $E_{8}$,
the fundamental lattice of the octonion structures in the
$D_{4}-D_{5}-E_{6}$ model described in
\href{http://xxx.lanl.gov/abs/hep-ph/9501252}{hep-ph/9501252}
\cite{SMI7}
and
\href{http://xxx.lanl.gov/abs/quant-ph/9503009}{quant-ph/9503009}.
\cite{SMI8}
\vspace{12pt}

The 240 nearest neighbors to the origin in the
$E_{8}$ lattice can be written in 7 different ways using
\newline
octonion coordinates with basis
\begin{equation}
\{ 1 ,e_{1},e_{2},e_{3},e_{4},e_{5},e_{6},e_{7} \}
\end{equation}
One way is:
\newline
16 vertices:
\begin{equation}
\pm 1, \pm e_{1}, \pm e_{2}, \pm e_{3}, \pm e_{4},
\pm e_{5}, \pm e_{6}, \pm e_{7}
\end{equation}
96 vertices:
\begin{equation}
\begin{array} {c}
(\pm 1 \pm e_{1} \pm e_{2} \pm e_{3}) / 2  \\
(\pm 1 \pm e_{2} \pm e_{5} \pm e_{7}) / 2  \\
(\pm 1 \pm e_{2} \pm e_{4} \pm e_{6}) / 2  \\
(\pm e_{4} \pm e_{5} \pm e_{6} \pm e_{7}) / 2  \\
(\pm e_{1} \pm e_{3} \pm e_{4} \pm e_{6}) / 2  \\
(\pm e_{1} \pm e_{3} \pm e_{5} \pm e_{7}) / 2  \\
\end{array}
\end{equation}
128 vertices:
\begin{equation}
\begin{array} {c}
(\pm 1 \pm e_{3} \pm e_{4} \pm e_{7}) / 2  \\
(\pm 1 \pm e_{1} \pm e_{5} \pm e_{6}) / 2  \\
(\pm 1 \pm e_{3} \pm e_{6} \pm e_{7}) / 2  \\
(\pm 1 \pm e_{1} \pm e_{4} \pm e_{7}) / 2  \\
(\pm e_{1} \pm e_{2} \pm e_{6} \pm e_{7}) / 2  \\
(\pm e_{2} \pm e_{3} \pm e_{4} \pm e_{7}) / 2  \\
(\pm e_{1} \pm e_{2} \pm e_{4} \pm e_{5}) / 2  \\
(\pm e_{2} \pm e_{3} \pm e_{5} \pm e_{6}) / 2  \\
\end{array}
\end{equation}

\vspace{12pt}

That the $E_{8}$ lattice is, in a sense,
fundamentally 4-dimensional
can be seen from several points of view:
\newline
\vspace{12pt}

the $E_{8}$ lattice nearest neighbor vertices have
only 4 non-zero coordinates,
like 4-dimensional spacetime with speed of light
$c$ = $\sqrt{3}$,
rather than 8 non-zero coordinates,
like 8-dimensional spacetime with speed of light $c$ = $\sqrt{7}$,
so the $E_{8}$ lattice light-cone structure appears to be
4-dimensional rather than 8-dimensional;
\newline
\vspace{12pt}

the representation of the $E_{8}$ lattice by quaternionic
icosians, as described by Conway and Sloane \cite{CON};
\newline
\vspace{12pt}

the Golden ratio construction of the $E_{8}$ lattice from
the $D_{4}$ lattice, which has a 24-cell nearest neighbor
polytope
(The construction starts with the 24 vertices of a 24-cell,
then adds Golden ratio points on each of the 96 edges of
the 24-cell, then extends the space to 8 dimensions
by considering the algebraicaly independent $\sqrt{5}$
part of the coordinates to be geometrically independent, and
finally doubling the resulting 120 vertices in 8-dimensional
space (by considering both the $D_{4}$ lattice and
its dual $D_{4}^{\ast}$)
to get the 240 vertices of the $E_{8}$ lattice nearest neighbor
polytope (the Witting polytope); and
\newline
\vspace{12pt}

the fact that the 240-vertex Witting polytope,
the $E_{8}$ lattice nearest neighbor polytope,
most naturally lives in 4 complex dimensions,
where it is self-dual, rather than in 8 real dimensions.
\newline
\vspace{12pt}

Some more material on such things can be found at
\newline
\href{http://www.gatech.edu/tsmith/home.html}{WWW
URL http://www.gatech.edu/tsmith/home.html} \cite{SMI6}.

\vspace{12pt}

In referring to Conway and Sloane \cite{CON}, bear in mind that
they use the convention (usual in working with lattices) that
the norm of a lattice distance is the square of the length of
the lattice distance.
\newline
\vspace{12pt}

\newpage

\subsection{4-dimensional HyperDiamond Lattice.}

The 4-dimensional HyperDiamond lattice $4HD$ is
\newline
$4HD$ = $D_{4} \cup  ([1] + D_{4})$.
\vspace{12pt}

The 4-dimensional HyperDiamond $4HD$ = $D_{4} \cup  ([1] + D_{4})$
is the ${\bf{Z}}^{4}$ hypercubic lattice with null edges.
\vspace{12pt}

It is the lattice that Michael Gibbs \cite{GIB} uses
in his Ph. D. thesis advised by David Finkelstein.
\vspace{12pt}

The 8 nearest neighbors to the origin in the
4-dimensional HyperDiamond $4HD$ lattice can be written
in octonion coordinates as:
\vspace{12pt}

\begin{equation}
\begin{array} {c}
(  1 + i + j + k) / 2  \\
(  1 + i - j - k) / 2  \\
(  1 - i + j - k) / 2  \\
(  1 - i - j + k) / 2  \\
(- 1 - i + j + k) / 2  \\
(- 1 + i - j + k) / 2  \\
(- 1 + i + j - k) / 2  \\
(- 1 - i - j - k) / 2  \\
\end{array}
\end{equation}

\vspace{12pt}

Here is an explicit construction of the 4-dimensional
HyperDiamond $4HD$ lattice nearest neighbors to the origin.
\vspace{12pt}

\newpage

START WITH THE 24 VERTICES OF A 24-CELL $D_{4}$:
\vspace{12pt}

\begin{equation}
\begin{array}{cccc}
+1   &    +1   &     0    &    0 \\
+1   &     0   &    +1    &    0 \\
+1   &     0   &     0    &   +1 \\
+1   &    -1   &     0    &    0 \\
+1   &     0   &    -1    &    0 \\
+1   &     0   &     0    &   -1 \\
-1   &    +1   &     0    &    0 \\
-1   &     0   &    +1    &    0 \\
-1   &     0   &     0    &   +1 \\
-1   &    -1   &     0    &    0 \\
-1   &     0   &    -1    &    0 \\
-1   &     0   &     0    &   -1 \\
 0   &    +1   &    +1    &    0 \\
 0   &    +1   &     0    &   +1 \\
 0   &    +1   &    -1    &    0 \\
 0   &    +1   &     0    &   -1 \\
 0   &    -1   &    +1    &    0 \\
 0   &    -1   &     0    &   +1 \\
 0   &    -1   &    -1    &    0 \\
 0   &    -1   &     0    &   -1 \\
 0   &     0   &    +1    &   +1 \\
 0   &     0   &    +1    &   -1 \\
 0   &     0   &    -1    &   +1 \\
 0   &     0   &    -1    &   -1 \\
\end{array}
\end{equation}

\vspace{12pt}

\newpage

SHIFT THE LATTICE BY A GLUE VECTOR,
\newline
BY ADDING
\vspace{12pt}

\begin{equation}
\begin{array}{cccc}
  0.5   &     0.5   &     0.5   &     0.5 \\
\end{array}
\end{equation}

\vspace{12pt}

TO GET 24 MORE VERTICES $[1] + D_{4}$:
\vspace{12pt}

\begin{equation}
\begin{array}{cccc}
+1.5   &    +1.5   &     0.5   &     0.5 \\
+1.5   &     0.5   &    +1.5   &     0.5 \\
+1.5   &     0.5   &     0.5   &    +1.5 \\
+1.5   &    -0.5   &     0.5   &     0.5 \\
+1.5   &     0.5   &    -0.5   &     0.5 \\
+1.5   &     0.5   &     0.5   &    -0.5 \\
-0.5   &    +1.5   &     0.5   &     0.5 \\
-0.5   &     0.5   &    +1.5   &     0.5 \\
-0.5   &     0.5   &     0.5   &    +1.5 \\
-0.5   &    -0.5   &     0.5   &     0.5 \\
-0.5   &     0.5   &    -0.5   &     0.5 \\
-0.5   &     0.5   &     0.5   &    -0.5 \\
 0.5   &    +1.5   &    +1.5   &     0.5 \\
 0.5   &    +1.5   &     0.5   &    +1.5 \\
 0.5   &    +1.5   &    -0.5   &     0.5 \\
 0.5   &    +1.5   &     0.5   &    -0.5 \\
 0.5   &    -0.5   &    +1.5   &     0.5 \\
 0.5   &    -0.5   &     0.5   &    +1.5 \\
 0.5   &    -0.5   &    -0.5   &     0.5 \\
 0.5   &    -0.5   &     0.5   &    -0.5 \\
 0.5   &     0.5   &    +1.5   &    +1.5 \\
 0.5   &     0.5   &    +1.5   &    -0.5 \\
 0.5   &     0.5   &    -0.5   &    +1.5 \\
 0.5   &     0.5   &    -0.5   &    -0.5 \\
\end{array}
\end{equation}

\vspace{12pt}

\newpage

FOR THE NEW COMBINED LATTICE $D_{4} \cup ([1] + D_{4})$,
\newline
THESE ARE 6 OF THE NEAREST NEIGHBORS
\newline
TO THE ORIGIN:
\vspace{12pt}

\begin{equation}
\begin{array}{cccc}
-0.5   &    -0.5   &     0.5   &     0.5 \\
-0.5   &     0.5   &    -0.5   &     0.5 \\
-0.5   &     0.5   &     0.5   &    -0.5 \\
 0.5   &    -0.5   &    -0.5   &     0.5 \\
 0.5   &    -0.5   &     0.5   &    -0.5 \\
 0.5   &     0.5   &    -0.5   &    -0.5 \\
\end{array}
\end{equation}

\vspace{12pt}

HERE ARE 2 MORE THAT COME FROM
\newline
ADDING THE GLUE VECTOR TO LATTICE VECTORS
\newline
THAT ARE NOT NEAREST NEIGHBORS OF THE ORIGIN:
\vspace{12pt}

\begin{equation}
\begin{array}{cccc}

 0.5   &     0.5   &     0.5   &     0.5 \\
-0.5   &    -0.5   &    -0.5   &    -0.5 \\
\end{array}
\end{equation}

\vspace{12pt}

THEY COME, RESPECTIVELY, FROM ADDING
\newline
THE GLUE VECTOR TO:
\vspace{12pt}

THE ORIGIN
\vspace{12pt}

\begin{equation}
\begin{array}{cccc}
  0   &     0   &     0   &     0 \\
\end{array}
\end{equation}

\vspace{12pt}

ITSELF;
\vspace{12pt}

AND
\vspace{12pt}

\newpage

THE LATTICE POINT
\vspace{12pt}

\begin{equation}
\begin{array}{cccc}
  -1   &     -1   &     -1   &     -1 \\
\end{array}
\end{equation}

\vspace{12pt}

WHICH IS SECOND ORDER, FROM
\vspace{12pt}

\begin{equation}
\begin{array}{cccc}
  -1   &     -1   &     0   &     0 \\
plus &&&  \\
  0   &     0   &     0   &     0 \\
\end{array}
\end{equation}

\vspace{12pt}

FROM
\vspace{12pt}

\begin{equation}
\begin{array}{cccc}
  -1   &     0   &     -1   &     0 \\
plus &&&  \\
  0   &     -1   &     0   &     -1 \\
\end{array}
\end{equation}

\vspace{12pt}

OR FROM
\vspace{12pt}

\begin{equation}
\begin{array}{cccc}
  -1   &     0   &     0   &     -1 \\
plus &&&  \\
  0   &     -1   &     -1   &     0 \\
\end{array}
\end{equation}

\vspace{12pt}

\newpage

\subsection{From 8 to 4 Dimensions.}

Dimensional reduction of the $8HD$ = $E_{8}$ lattice spacetime
to 4-dimensional spacetime reduces each of the $D_{8}$
lattices in the
$$E_{8} = 8HD = D_{8} \cup  ([1] + D_{8})$$
lattice to $D_{4}$ lattices.
\vspace{12pt}

Therefore, we should get a 4-dimensional HyperDiamond
$$4HD = D_{4} \cup  ([1] + D_{4})$$
lattice.
\vspace{12pt}

To see this, start with $E_{8} = 8HD = D_{8} \cup  ([1] + D_{8})$.
\vspace{12pt}

We can write:
\vspace{12pt}

\begin{equation}
D_{8} = \{ (D_{4},0,0,0,0) \} \cup \{ (0,0,0,0,D_{4}) \}
\cup \{ (1,0,0,0,1,0,0,0) + (D_{4},D_{4}) \}
\end{equation}

\vspace{12pt}

The third term is the diagonal term of
\newline
an orthogonal decomposition of $D_{8}$, and
\vspace{12pt}

the first two terms are the orthogonal
\newline
associative physical 4-dimensional spacetime
\newline
and
\newline
coassociative 4-dimensional internal symmetry space
\newline
as described in
\newline
{\it Standard Model plus Gravity from
\newline
Octonion Creators
and Annihilators},
\newline
\href{http://xxx.lanl.gov/abs/quant-th/9503009}{quant-th/9503009}
\cite{SMI8}.
\vspace{12pt}

Now, we see that the orthogonal decomposition of
8-dimensional spacetime into 4-dimensional associative
physical spacetime plus 4-dimensional internal symmetry space
gives a decomposition of $D_{8}$ into $D_{4} \oplus D_{4}$.
\vspace{12pt}

\newpage

Since $E_{8} = D_{8} \cup  ([1] + D_{8})$, and
\newline
since $[1] = (0.5,0.5,0.5,0.5,0.5,0.5,0.5,0.5)$ can
\newline
be decomposed by

\begin{equation}
[1] = (0.5,0.5,0.5,0.5,0,0,0,0) \oplus (0,0,0,0,0.5,0.5,0.5,0.5)
\end{equation}

we have
\vspace{12pt}

\begin{eqnarray}
E_{8} & = & D_{8} \cup  ([1] + D_{8}) \\
 & & \nonumber \\
 & = & ((D_{4},0,0,0,0) \oplus (0,0,0,0,D_{4})) \nonumber \\
 & & \cup \nonumber \\
 & & (((0.5,0.5,0.5,0.5,0,0,0,0) \oplus (0,0,0,0,0.5,0.5,0.5,0.5))
\nonumber \\
 & & + \nonumber \\
 & & ((D_{4},0,0,0,0) \oplus (0,0,0,0,D_{4}))) \nonumber \\
 & & \nonumber \\
 & = & ((D_{4},0,0,0,0) \cup ((0.5,0.5,0.5,0.5,0,0,0,0) +
(D_{4},0,0,0,0))) \nonumber \\
 & & \oplus \nonumber \\
 & & ((0,0,0,0,D_{4}) \cup ((0,0,0,0,0.5,0.5,0.5,0.5) +
(0,0,0,0,D_{4}))) \nonumber
\end{eqnarray}

\vspace{12pt}

Since $4HD$ is $D_{4} \cup ([1] + D_{4})$,

\begin{equation}
E_{8} = 8HD = 4HD_{a} \oplus 4HD_{ca}
\end{equation}

where $4HD_{a}$ is the 4-dimensional associative physical
spacetime and
\newline
$4HD_{ca}$ is the 4-dimensional coassociative internal symmetry space.
\vspace{12pt}

\newpage

\section{Internal Symmetry Space.}

$4HD_{ca}$ is the 4-dimensional coassociative Internal Symmetry Space
of the 4-dimensional HyperDiamond Feynman checkerboard version of
the $D_{4}-D_{5}-E_{6}$ model.
\vspace{12pt}

Physically, the $4HD_{ca}$ Internal Space should be thought of
as a space "inside" each vertex of the $4HD_{a}$ Feynman
checkerboard spacetime, sort of like a Kaluza-Klein structure.
\vspace{12pt}

The 4 dimensions of the $4HD_{ca}$ Internal Symmetry Space are:
\newline
electric charge;
\newline
red color charge;
\newline
green color charge; and
\newline
blue color charge.
\vspace{12pt}

Each vertex of the $4HD_{ca}$ lattice has 8 nearest neighbors,
connected by lightcone links.  They have the algebraic structure
of the 8-element quaternion group $<2,2,2>$. \cite{COX2}
\vspace{12pt}

Each vertex of the $4HD_{ca}$ lattice has 24 next-to-nearest neighbors,
connected by two lightcone links.  They have the algebraic structure
of the 24-element binary tetrahedral group $<3,3,2>$ that is
associated with the 24-cell and the $D_{4}$ lattice. \cite{COX2}
\vspace{12pt}

\newpage

The 1-time and 3-space dimensions of the $4HD_{a}$ spacetime
can be represented by the 4 future lightcone links and the
4 past lightcone links as in the following pair of "Square Diagrams"
of the 4 lines connecting the future ends
of the 4 future lightcone links
and
of the 4 lines connecting the past ends
of the 4 past lightcone links:

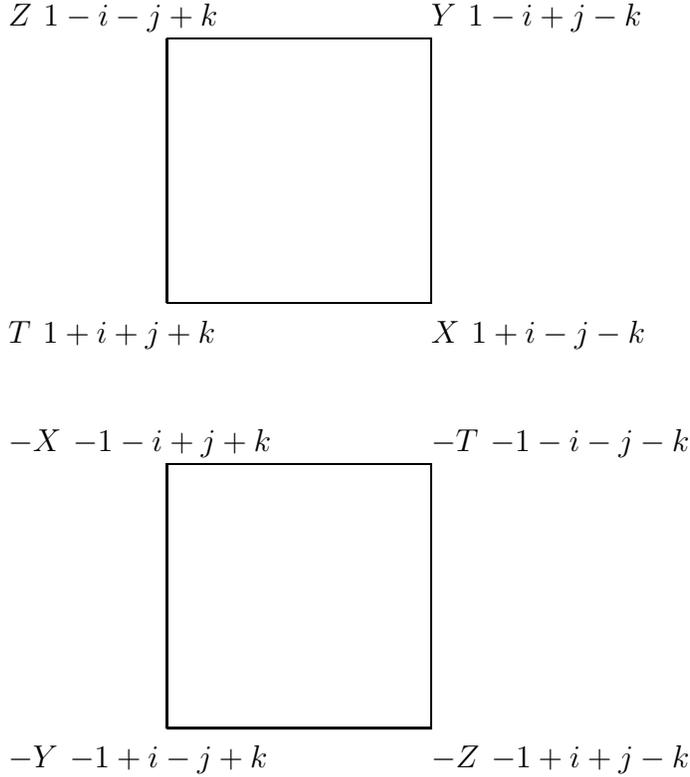
\begin{picture}(140,140)

\put(65,15){\line(1,0){100}}
\put(165,15){\line(0,1){100}}
\put(165,115){\line(-1,0){100}}
\put(65,115){\line(0,-1){100}}
\put(5,0){{\bf $T$ }$1+i+j+k$}
\put(165,0){{\bf $X$ }$1+i-j-k$}
\put(165,120){{\bf $Y$ }$1-i+j-k$}
\put(5,120){{\bf $Z$ }$1-i-j+k$}

\end{picture}

\begin{picture}(160,160)

\put(65,15){\line(1,0){100}}
\put(165,15){\line(0,1){100}}
\put(165,115){\line(-1,0){100}}
\put(65,115){\line(0,-1){100}}
\put(5,0){{\bf $-Y$ }$-1+i-j+k$}
\put(165,0){{\bf $-Z$ }$-1+i+j-k$}
\put(165,120){{\bf $-T$ }$-1-i-j-k$}
\put(5,120){{\bf $-X$ }$-1-i+j+k$}

\end{picture}

\vspace{12pt}

The 8 links $\{ {\bf{T,X,Y,Z,-T,-X,-Y,-Z}} \}$ correspond
to the 8 root vectors of the $Spin(5)$ de Sitter gravitation
gauge group, which has an 8-element Weyl group $S_{2}^{2} \times S_{2}$.
\newline
The symmetry group of the 4 links of the future lightcone is $S_{4}$,
the Weyl group of the 15-dimensional Conformal group $SU(4)$ = $Spin(6)$.
\newline
10 of the 15 dimensions make up the de Sitter $Spin(5)$ subgroup, and
\newline
the other 5 fix the "symmetry-breaking direction" and scale of the
Higgs mechanism (for more on this, see \cite{SMI7} and
\newline
\href{http://www.gatech.edu/tsmith/cnfGrHg.html}{WWW
URL http://www.gatech.edu/tsmith/cnfGrHg.html}).

\newpage

Similarly, the E-electric and RGB-color dimensions of the $4HD_{ca}$
Internal Symmetry Space can be represented by
the 4 future lightcone links and the
4 past lightcone links.
\vspace{12pt}

However, in the $4HD_{ca}$ Internal Symmetry Space the
E Electric Charge should be treated as independent of
the RGB Color Charges.
\vspace{12pt}

As a result the following pair of "Square Diagrams"
look more like "Triangle plus Point Diagrams".

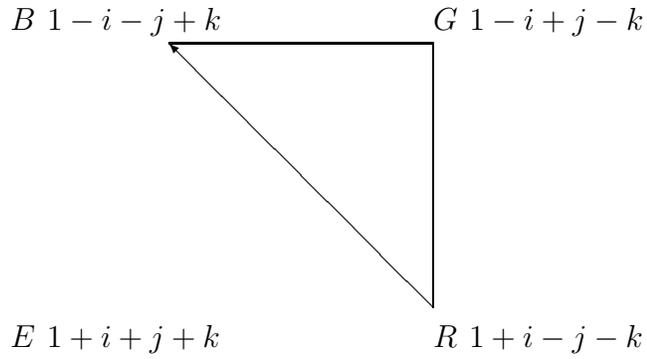
\begin{picture}(200,200)

\put(165,15){\line(0,1){100}}
\put(165,115){\line(-1,0){100}}
\put(5,0){{\bf $E$ }$1+i+j+k$}
\put(165,0){{\bf $R$ }$1+i-j-k$}
\put(165,120){{\bf $G$ }$1-i+j-k$}
\put(5,120){{\bf $B$ }$1-i-j+k$}
\put(165,15){\vector(-1,1){100}}

\end{picture}

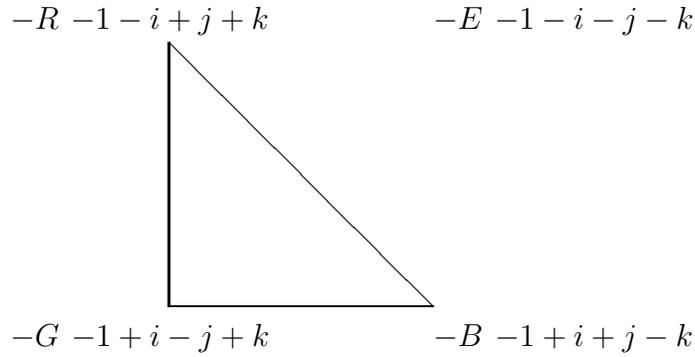
\begin{picture}(200,200)

\put(65,15){\line(1,0){100}}
\put(65,115){\line(0,-1){100}}
\put(5,0){{\bf $-G$ }$-1+i-j+k$}
\put(165,0){{\bf $-B$ }$-1+i+j-k$}
\put(165,120){{\bf $-E$ }$-1-i-j-k$}
\put(5,120){{\bf $-R$ }$-1-i+j+k$}
\put(165,15){\line(-1,1){100}}

\end{picture}

The 2+6 links $\{ {\bf{E,-E;R,G,B,-R,-G,-B}} \}$ correspond to:
\vspace{12pt}

the 2 root vectors of the weak force $SU(2)$,
which has a 2-element Weyl group $S_{2}$; and
\vspace{12pt}

the 6 root vectors of the color force $SU(3)$,
which has a 6-element Weyl group $S_{3}$.
\vspace{12pt}

\newpage

\section{Protons, Pions, and Physical Gravitons.}

In his 1994 Georgia Tech Ph. D. thesis under David Finkelstein,
{\it{Spacetime as a Quantum Graph}}, Michael Gibbs \cite{GIB}
describes some 4-dimensional HyperDiamond lattice structures,
that he considers likely candidates to represent physical
particles.
\vspace{12pt}

The terminology used by Michael Gibbs in his thesis \cite{GIB}
is useful with respect to the model he constructs.  Since his
model is substantially different from my HyperDiamond Feynman
checkerboard in some respects, I use a different terminology
here.  However, I want to make it clear that I have borrowed
these particular structures from his thesis.
\vspace{12pt}

Three useful HyperDiamond structures are:
\vspace{12pt}

3-link Rotating Propagator, useful for building a
proton out of 3 quarks;
\vspace{12pt}

2-link Exchange Propagator, useful for building a
pion out of a quark and an antiquark; and
\vspace{12pt}

4-link Propagator, useful for building a physical spin-2
physical graviton out of $Spin(5)$ Gauge bosons..
\vspace{12pt}

In the 2-dimensional Feynman checkerboard, there is
only one massive particle, the electron.
\vspace{12pt}

What about the $D_{4}-D_{5}-E_{6}$ model, or any other
model that has different particles with different masses?
\vspace{12pt}

In the context of Feynman checkerboards, mass is just
the amplitude for a particle to have a change of direction
in its path.
\vspace{12pt}

More massive particles will change direction more often.
\vspace{12pt}

In the $D_{4}-D_{5}-E_{6}$ model, the $4HD$ Feynman checkerboard
fundamental path segment length $\epsilon$ of any particle
the Planck length $L_{PL}$.
\vspace{12pt}

However, in the sum over paths for a particle of mass $m$,
\newline
it is a useful approximation to consider the path segment
length to be the Compton wavelength $L_{m}$ of the mass $m$,
$$L_{m} =  h/mc$$
\vspace{12pt}

That is because the distances between direction changes in the vast
bulk of the paths will be at least $L_{m}$, and
\newline
those distances will be approximately integral multiples of $L_{m}$,
\newline
so that $L_{m}$ can be used as the effective path segment length.
\vspace{12pt}

This is an important approximation because the Planck length $L_{PL}$
is about $10^{-33}$ cm,
\newline
while the effective length $L_{100GeV}$ for a
particle of mass $100$ $GeV$ is about $10^{-16}$ cm.
\vspace{12pt}

In this section, the $4HD$ lattice is given quaternionic coordinates.
\newline
The orgin $0$ designates the beginning of the path.
\newline
The 4 future lightcone links from the origin are given the coordinates
\newline
$1+i+j+k$, $1+i-j-k$, $1-i+j-k$, $1-i-j+k$
\vspace{12pt}

The path of a "Particle" at rest in space, moving 7 steps in time,
is denoted by
\vspace{12pt}

\[ \begin{array}{|c|}
Particle \\
0 \\
1 \\
2 \\
3 \\
4 \\
5 \\
6 \\
7
\end{array} \]

Note that since the $4HD$ speed of light is $\sqrt{3}$, the
path length is $7 \sqrt{3}$.
\vspace{12pt}

\newpage

The path of a "Particle" moving along a lightcone path in
the $1+i+j+k$ direction for 7 steps with no change of direction
is
\vspace{12pt}

\[ \begin{array}{|c|}
Particle \\
0 \\
 1+i+j+k \\
2+2i+2j+2k \\
3+3i+3j+3k \\
4+4i+4j+4k \\
5+5i+5j+5k \\
6+6i+6j+6k \\
7+7i+7j+7k
\end{array} \]

\vspace{12pt}

At each step in either path, the future lightcone can
be represented by a "Square Diagram" of lines connecting the future
ends of the 4 future lightcone links leading from the vertex at
which the step begins.

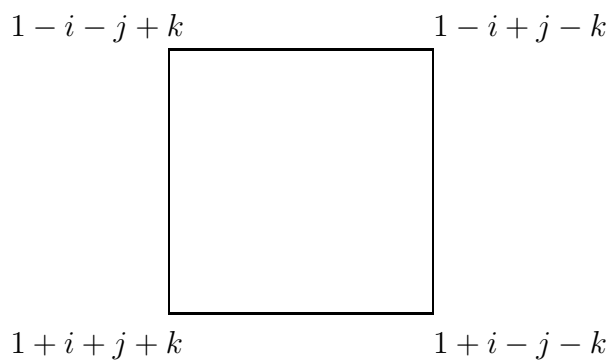
\begin{picture}(200,200)

\put(65,15){\line(1,0){100}}
\put(165,15){\line(0,1){100}}
\put(165,115){\line(-1,0){100}}
\put(65,115){\line(0,-1){100}}
\put(5,0){$1+i+j+k$}
\put(165,0){$1+i-j-k$}
\put(165,120){$1-i+j-k$}
\put(5,120){$1-i-j+k$}

\end{picture}

\vspace{12pt}

In the following subsections, protons, pions, and physical gravitons
will be represented by multiparticle paths.
\newline
The multiple particles representing protons, pions, and physical
gravitons will be shown on sequences of such Square Diagrams,
as well as by a sequence of coordinates.
\newline
The coordinate sequences will be given only for a representative
sequence of timelike steps, with no space movement, because
\newline
the notation for a timelike sequence is clearer and
\newline
it is easy to transform a sequence of timelike steps into
a sequence of lightcone link steps, as shown above.
\vspace{12pt}

Only in the case of gravitons will it be useful to explicitly
discuss a path that moves in space as well as time.
\vspace{12pt}

\newpage

\subsection{3-Quark Protons.}

In this $4HD$ Feynman checkerboard version of the $D_{4}-D_{5}-E_{6}$
model, protons are made up of 3 valence first generation quarks,
(two up quarks and one down quark), with one Red, one Green,
and one Blue in color.
\newline
The proton bound state of 3 valence
quarks has a soliton structure (see
\href{http://www.gatech.edu/tsmith/SolProton.html}{WWW
URL http://www.gatech.edu/tsmith/SolProton.html} \cite{SMI6}).
\vspace{12pt}

The complicated structure of the sea quarks and binding gluons
can be ignored in a $4HD$ Feynman checkerboard approximation that
uses the R, G, and B valence quarks and path length $L_{313GeV}$
that is the Compton wavelength of the first generation quark
constituent mass.
\vspace{12pt}

The $4HD$ structure used to approximate the proton is the
3-link Rotating Propagator of the thesis of Michael Gibbs \cite{GIB}.
\vspace{12pt}

\newpage

Here is a coordinate sequence representation of the approximate
$4HD$ Feynman checkerboard path of a proton:

\vspace{12pt}

\[ \begin{array}{|c|c|c|}
R-Quark & G-Quark & B-Quark \\
1+i-j-k & 1-i+j-k & 1-i-j+k \\
2-i+j-k & 2-i-j+k & 2+i-j-k \\
3-i-j+k & 3+i-j-k & 2-i+j-k \\
4+i-j-k & 4-i+j-k & 4-i-j+k \\
5-i+j-k & 5-i-j+k & 5+i-j-k \\
6-i-j+k & 6+i-j-k & 6-i+j-k \\
7+i-j-k & 7-i+j-k & 7-i-j+k
\end{array} \]

\vspace{12pt}

The following page contains a Square Diagram representation of
the approximate $4HD$ Feynman checkerboard path of a proton:
\vspace{12pt}

\newpage

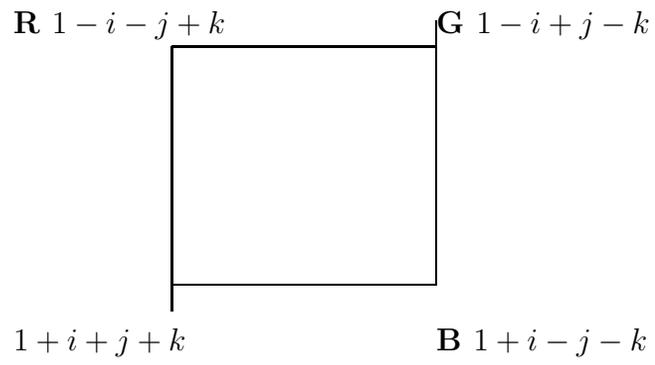
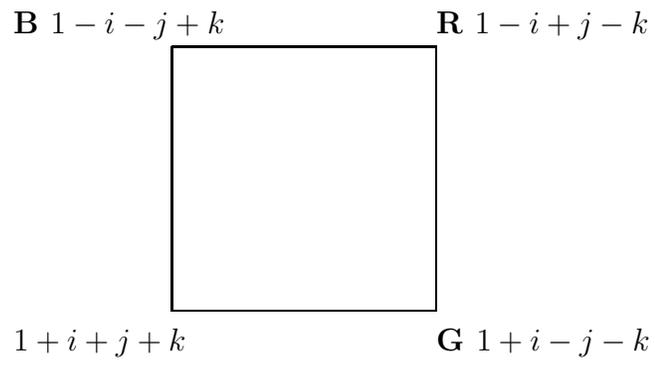
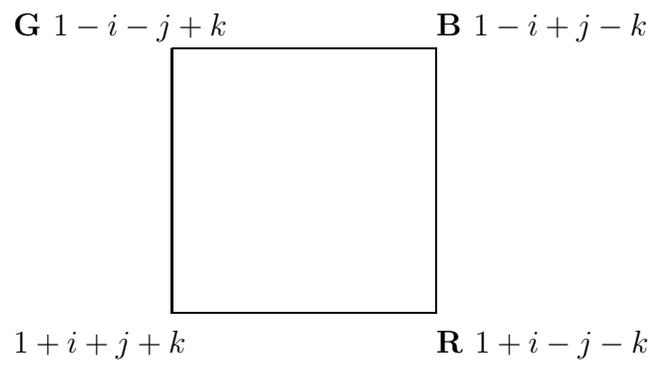
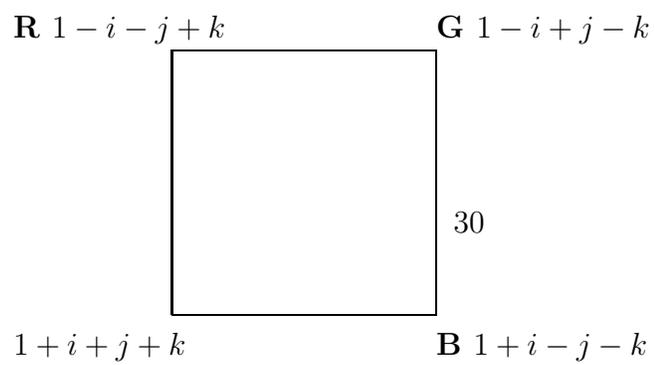
\begin{picture}(200,200)(0,410)

\put(65,595){\line(1,0){100}}
\put(165,595){\line(0,1){100}}
\put(165,685){\line(-1,0){100}}
\put(65,685){\line(0,-1){100}}
\put(5,570){$1+i+j+k$}
\put(165,570){{\bf B }$1+i-j-k$}
\put(165,690){{\bf G }$1-i+j-k$}
\put(5,690){{\bf R }$1-i-j+k$}

\put(65,395){\line(1,0){100}}
\put(165,395){\line(0,1){100}}
\put(165,495){\line(-1,0){100}}
\put(65,495){\line(0,-1){100}}
\put(5,380){$1+i+j+k$}
\put(165,380){{\bf G }$1+i-j-k$}
\put(165,500){{\bf R }$1-i+j-k$}
\put(5,500){{\bf B }$1-i-j+k$}

\put(65,205){\line(1,0){100}}
\put(165,205){\line(0,1){100}}
\put(165,305){\line(-1,0){100}}
\put(65,305){\line(0,-1){100}}
\put(5,190){$1+i+j+k$}
\put(165,190){{\bf R }$1+i-j-k$}
\put(165,310){{\bf B }$1-i+j-k$}
\put(5,310){{\bf G }$1-i-j+k$}

\put(65,15){\line(1,0){100}}
\put(165,15){\line(0,1){100}}
\put(165,115){\line(-1,0){100}}
\put(65,115){\line(0,-1){100}}
\put(5,0){$1+i+j+k$}
\put(165,0){{\bf B }$1+i-j-k$}
\put(165,120){{\bf G }$1-i+j-k$}
\put(5,120){{\bf R }$1-i-j+k$}

\end{picture}

\newpage

\subsection{Quark-AntiQuark Pions.}

In this $4HD$ Feynman checkerboard version of the $D_{4}-D_{5}-E_{6}$
model, pions are made up of first generation valence Quark-AntiQuark
pairs.
\newline
The pion bound state of valence Quark-AntiQuark pairs has a soliton
structure that would, projected onto a 2-dimensional spacetime, be
a Sine-Gordon breather (see
\href{http://www.gatech.edu/tsmith/SnGdnPion.html}{WWW
URL http://www.gatech.edu/tsmith/SnGdnPion.html} \cite{SMI6}).
\vspace{12pt}

The complicated structure of the sea quarks and binding gluons
can be ignored in a $4HD$ Feynman checkerboard approximation that
uses the valence Quark and AntiQuark and path length $L_{313GeV}$
that is the Compton wavelength of the first generation quark
constituent mass.
\vspace{12pt}

The $4HD$ structure used to approximate the pion is the
2-link Exchange Propagator.
\vspace{12pt}

\newpage

Here is a coordinate sequence representation of the approximate
$4HD$ Feynman checkerboard path of a pion:
\vspace{12pt}

\[ \begin{array}{|c|c|}
Quark & AntiQuark \\

0 & 0 \\
1+i+j+k & 1-i+j-k \\
2 & 2 \\
3+i-j-k & 3-i-j+k \\
4 & 4 \\
5-i+j-k & 5+i+j+k \\
6 & 6 \\
7-i-j+k & 7+i-j-k \\
8 & 8 \\
9+i+j+k & 9-i+j-k \\
10 & 10 \\
11+i-j-k & 11-i-j+k \\
12 & 12 \\
13-i+j-k & 13+i+j+k \\
14 & 14 \\
15-i-j+k & 15+i-j-k \\
16 & 16
\end{array} \]

\vspace{12pt}

The following page contains a Square Diagram representation of
the approximate $4HD$ Feynman checkerboard path of a pion:
\vspace{12pt}

\newpage

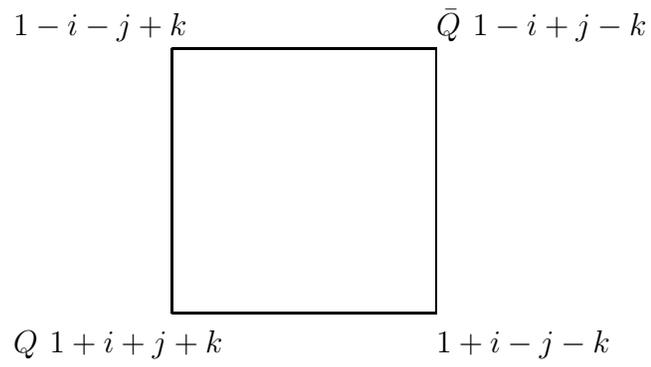
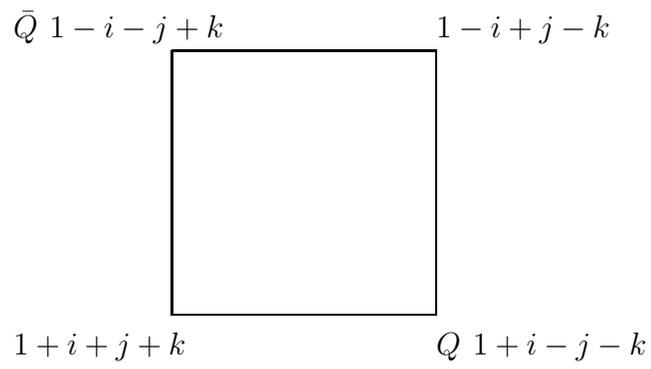
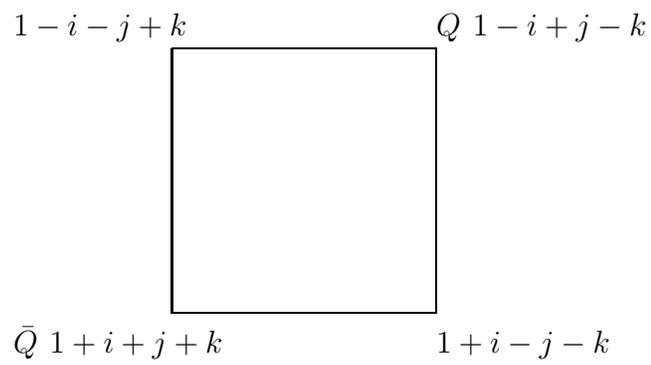
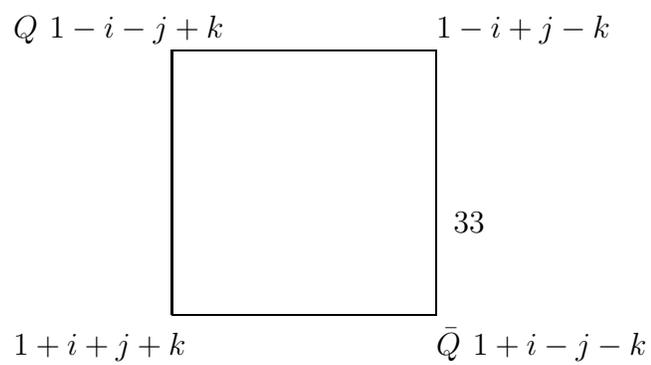
\begin{picture}(200,200)(0,410)

\put(65,585){\line(1,0){100}}
\put(165,585){\line(0,1){100}}
\put(165,685){\line(-1,0){100}}
\put(65,685){\line(0,-1){100}}
\put(5,570){{\bf $Q$ }$1+i+j+k$}
\put(165,570){$1+i-j-k$}
\put(165,690){{\bf $\bar{Q}$ }$1-i+j-k$}
\put(5,690){$1-i-j+k$}

\put(65,395){\line(1,0){100}}
\put(165,395){\line(0,1){100}}
\put(165,495){\line(-1,0){100}}
\put(65,495){\line(0,-1){100}}
\put(5,380){$1+i+j+k$}
\put(165,380){{\bf $Q$ }$1+i-j-k$}
\put(165,500){$1-i+j-k$}
\put(5,500){{\bf $\bar{Q}$ }$1-i-j+k$}

\put(65,205){\line(1,0){100}}
\put(165,205){\line(0,1){100}}
\put(165,305){\line(-1,0){100}}
\put(65,305){\line(0,-1){100}}
\put(5,190){{\bf $\bar{Q}$ }$1+i+j+k$}
\put(165,190){$1+i-j-k$}
\put(165,310){{\bf $Q$ }$1-i+j-k$}
\put(5,310){$1-i-j+k$}

\put(65,15){\line(1,0){100}}
\put(165,15){\line(0,1){100}}
\put(165,115){\line(-1,0){100}}
\put(65,115){\line(0,-1){100}}
\put(5,0){$1+i+j+k$}
\put(165,0){{\bf $\bar{Q}$ }$1+i-j-k$}
\put(165,120){$1-i+j-k$}
\put(5,120){{\bf $Q$ }$1-i-j+k$}

\end{picture}

\newpage

\subsection{Spin-2 Physical Gravitons.}

In this $4HD$ Feynman checkerboard version of the $D_{4}-D_{5}-E_{6}$
model, spin-2 physical gravitons are made up of the 4 translation
spin-1 gauge bosons of the 10-dimensional $Spin(5)$ de Sitter subgroup
of the 15-dimensional $Spin(6)$ Conformal group used to
construct Einstein-Hilbert gravity in the $D_{4}-D_{5}-E_{6}$ model
described in
\href{http://xxx.lanl.gov/abs/hep-ph/9501252}{hep-ph/9501252}
and
\href{http://xxx.lanl.gov/abs/quant-ph/9503009}{quant-ph/9503009}.
\vspace{12pt}

These spin-2 physical gravitons are massless, but they can have
energy up to and including the Planck mass.
\vspace{12pt}

The Planck energy spin-2 physical gravitons are really
fundamental structures with $4HD$ Feynman checkerboard
path length $L_{Planck}$.
\vspace{12pt}

\newpage

Here is a coordinate sequence representation of
\newline
the $4HD$ Feynman checkerboard path of a fundamental
\newline
Planck-mass spin-2 physical graviton,
\newline
where ${\bf T,  X, Y, Z }$ represent the 4 translation
\newline
generator of the $Spin(5)$ de Sitter group:
\vspace{12pt}

\[ \begin{array}{|c|c|c|c|}
{\bf T } & {\bf X } & {\bf Y } & {\bf Z } \\
0 & 0 & 0 & 0 \\
1+i+j+k & 1+i-j-k & 1-i+j-k & 1-i-j+k \\
2 & 2 & 2 & 2 \\
3+i+j+k & 3+i-j-k & 3-i+j-k & 3-i-j+k \\
4 & 4 & 4 & 4 \\
5+i+j+k & 5+i-j-k & 5-i+j-k & 5-i-j+k \\
6 & 6 & 6 & 6 \\
7+i+j+k & 7+i-j-k & 7-i+j-k & 7-i-j+k \\
8 & 8 & 8 & 8
\end{array} \]

\vspace{12pt}

\vspace{12pt}

The following is a Square Diagram representation of the
$4HD$ Feynman checkerboard path of a fundamental Planck-mass
spin-2 physical graviton:
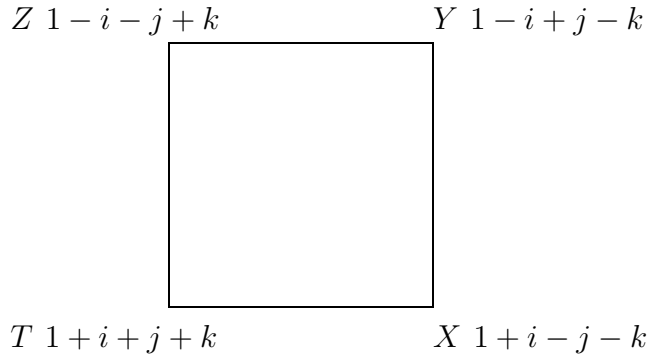
\vspace{12pt}

\begin{picture}(200,200)

\put(65,15){\line(1,0){100}}
\put(165,15){\line(0,1){100}}
\put(165,115){\line(-1,0){100}}
\put(65,115){\line(0,-1){100}}
\put(5,0){{\bf $T$ }$1+i+j+k$}
\put(165,0){{\bf $X$ }$1+i-j-k$}
\put(165,120){{\bf $Y$ }$1-i+j-k$}
\put(5,120){{\bf $Z$ }$1-i-j+k$}

\end{picture}

\vspace{12pt}

The representation above is for a timelike path at rest in space.
\vspace{12pt}

\newpage

With respect to gravitons, we can see something new and different
by letting the path move in space as well.
\vspace{12pt}

Let ${\bf T}$ and ${\bf Y}$ represent time and longitudinal space, and
\newline
${\bf X}$ and  ${\bf Z}$ represent transverse space.
\vspace{12pt}

Then, as discussed in Feynman's {\it Lectures on Gravitation},
pp. 41-42 \cite{FEY}, the Square Diagram representation shows
that our spin-2 physical graviton is indeed a spin-2 particle.
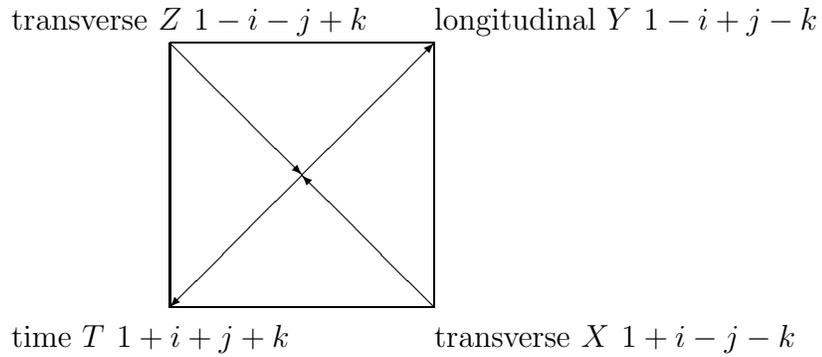
\vspace{12pt}

\begin{picture}(200,200)

\put(65,15){\line(1,0){100}}
\put(165,15){\line(0,1){100}}
\put(165,115){\line(-1,0){100}}
\put(65,115){\line(0,-1){100}}
\put(5,0){time {\bf $T$ }$1+i+j+k$}
\put(165,0){transverse {\bf $X$ }$1+i-j-k$}
\put(165,120){longitudinal {\bf $Y$ }$1-i+j-k$}
\put(5,120){transverse {\bf $Z$ }$1-i-j+k$}
\put(115,65){\vector(1,1){50}}
\put(115,65){\vector(-1,-1){50}}
\put(165,15){\vector(-1,1){50}}
\put(65,115){\vector(1,-1){50}}

\end{picture}

\vspace{12pt}

Spin-2 physical gravitons of energy less than the Planck mass
are more complicated composite gauge boson structures with
approximate $4HD$ Feynman checkerboard path length
$L_{graviton energy}$.
\vspace{12pt}

They can be deformed from a square shape, but retain
their spin-2 nature as described by Feynman \cite{FEY}.


\vspace{12pt}


\newpage



\begin{thebibliography}{99}


\bibitem{CON} J. Conway and N. Sloane, {\it Sphere Packings,
Lattices, and Groups, 2nd ed}, Springer-Verlag (1993).

\bibitem{COX2} H. Coxeter,
{\it Regular Complex Polytopes, 2nd ed},
Cambridge (1991).

\bibitem{DIX4} G. Dixon, {\it Division algebras:  octonions,
quaternions, complex numbers, and the algebraic design of physics},
Kluwer (1994).

\bibitem{DIX5} G. Dixon, {\it Octonion X-product Orbits},
\href{http://xxx.lanl.gov/abs/hep-th/9410202}{hep-th/9410202}.

\bibitem{DIX6} G. Dixon,
{\it Octonion X-product and
\newline
Octonion $E_{8}$ Lattices},
\href{http://xxx.lanl.gov/abs/hep-th/9411063}{hep-th/9411063}.

\bibitem{DIX7} G. Dixon, {\it Octonions:  $E_{8}$ Lattice
to ${\Lambda}_{16}$},
\href{http://xxx.lanl.gov/abs/hep-th/9501007}{hep-th/9501007}.

\bibitem{FEY} R. Feynman, {\it Lectures on Gravitation, 1962-63},
Caltech (1971).

\bibitem{FEY1} R. Feynman, {\it Notes on the Dirac Equation},
Caltech archives, box 13, folder 6 (1946).

\bibitem{FEY2} R. Feynman and Hibbs, {\it  Quantum Mechanics and
Path Integrals}, McGraw-Hill (1965).

\bibitem{GER} H. Gersch, Int. J. Theor. Phys. 20 (1981) 491.

\bibitem{GIB} M. Gibbs, {\it Spacetime as a Quantum Graph},
Ph. D. thesis of J. Michael Gibbs,
with advisor David Finkelstein, Georgia Tech (1994).

\bibitem{HRG} H. Hrgovcic, {\it Quantum mechanics on space-time
lattices using path integrals in a Minkowski metric},
MIT Ph. D. thesis under T. Toffoli, July 1992.

\bibitem{SMI6} F. Smith,
\href{http://www.gatech.edu/tsmith/home.html}{WWW URL
http://www.gatech.edu/tsmith/home.html}.

\bibitem{SMI7} F. Smith,  {\it Gravity and the Standard Model
with 130 GeV Truth Quark from $D_{4}-D_{5}-E_{6}$ Model using
$3 \times 3$ Octonion Matrices}, preprint: THEP-95-1;
\href{http://xxx.lanl.gov/abs/hep-ph/9501252}{hep-ph/9501252}.

\bibitem{SMI8} F. Smith,  {\it Standard Model plus Gravity
from Octonion Creators and Annihilators}, preprint: THEP-95-2;
\href{http://xxx.lanl.gov/abs/quant-th/9503009}{quant-th/9503009}.


\end{thebibliography}
\end{document}